\documentclass[preprint,3p,twocolumn]{elsarticle}
\usepackage{color}
\usepackage{amsmath}
\usepackage{amssymb}
\usepackage{graphicx}
\usepackage[colorlinks=true,linkcolor=red]{hyperref}

\journal{Physics Letters A}
\biboptions{square,comma}

\begin{document}
\begin{frontmatter}

\title{Discreteness effects on the fluxon interaction with the dipole
impurity in the Josephson transmission line}
\author[is]{Ivan O. Starodub}
\ead{starodub@bitp.kiev.ua}

\author[yz]{Yaroslav Zolotaryuk\corref{cor1}}
\cortext[cor1]{Corresponding author}
\ead{yzolo@bitp.kiev.ua}
\address
{Bogolyubov Institute for Theoretical Physics, National Academy of
Sciences of Ukraine, vul. Metrologichna 14-b, Kyiv 03143, Ukraine}
\date{\today}

\begin{abstract}
The influence of  discreteness on the fluxon
scattering on the dipole-like impurity is studied. This
kind of impurity is used to model the  qubit inductively
coupled to the Josephson transmission line (JTL). The previously proposed 
fluxon assisted readout process of the qubit state is based on measuring
the passage time through the dipole impurity. 
The aim of this work is to clarify the role of the discreteness in
this qubit readout process.
It is demonstrated that the fluxon delay time on the qubit
impurity increases significantly if the discreteness of the
JTL increases. Also it is shown that the difference between the
fluxon threshold currents for the positive and negative qubits
decreases with the increase of discreteness. 
\end{abstract}


\end{frontmatter}

\section{Introduction}

The Josephson effect is among the the most remarkable phenomena in
physics of superconductors \cite{barone82,kkl86}. Spatially extended 
systems that consist of Josephson junctions (JJs) have been
an important area of scientific research during the last decades.
One of the reasons for this activity is the abundance of various 
nontrivial nonlinear
phenomena observed in such systems and their applications 
such as topological soliton
(fluxon) \cite{fd73ssc,ddkp85prl,u98pd} and discrete breather \cite{tmo00prl,baufz00prl}
observation, dynamical chaos \cite{b-jgif82prl,k96rpp}, metastructures \cite{lt13sst}, metrological applications \cite{bkkmp12mst} to name a few.

One of the important application of fluxons in Josephson junctions
is quantum computation \cite{kwu02pssb}-\cite{ow23prapp}. 
The qubit readout process suggested in
\cite{ars06prb} and studied further both analytically and experimentally 
in \cite{fssk-s07prb}-\cite{skprik15prb},\cite{sz19pla} is based on the Josephson transmission line inductively coupled to a qubit. The fluxon (Josephson vortex) is launched 
at the JTL end, and, since its
interaction with the $|0\rangle$ and $|1\rangle$ qubit states takes different times,
its arrival at the opposite JTL end occurs after different time intervals.
Hence, by measuring the delay time one can determine the state of the qubit. 

It should be noted that although the JTL are discrete, all the theoretical
research on the fluxon qubit readout \cite{ars06prb,fssk-s07prb,skprik15prb,sz19pla} 
has been performed in the continuum approximation. Therefore it is natural 
to investigate the role of discreteness in the qubit readout process in JTLs.
Topological soliton interaction with spatial inhomogeneities 
has been studied rather well \cite{fpe89pla,bk91prb}, however, the case of an dipole 
inhomogeneity in the discrete media requires much more attention.

The main goal of this work is to investigate the effect of discreteness
on the qubit readout and to find a way to maximize the sensitivity of the qubit readout process.

This paper is organized as follows. In the next Section we 
describe the model. In Sec. \ref{sec3} the main results on the fluxon
transmission are presented and the threshold current is computed.
In Sec. \ref{sec4} the fluxon delay time on the qubit is obtained.
Discussion and conclusions are given in the last Section.

\section{The model}
\label{model}

We consider the Josephson junction array or, in other words, the
Josephson transmission line (JTL) and a qubit coupled inductively
to it. The Hamiltonian of the JTL interacting with the qubit
can be written \cite{ars06prb,fssk-s07prb} as follows:
\begin{eqnarray}\nonumber
H&=&\sum_{n=1}^N \left [\frac{Q_n^2}{2C}+E_J (1-\cos \phi_n)+
\right.\\
 &+& \left.  
\frac{1}{2L}\left(\frac{\Phi_0}{2\pi} \right)^2(\phi_{n+1}-
\phi_n-\phi_n^{(e)})^2 \right ].
\label{1}
\end{eqnarray}
Here the charge on the $n$th junction equals $Q_n=C\hbar\dot{\phi}_n/(2e)$, the Josephson coupling energy $E_J=\hbar I_c/(2e)$. Each junction
is described by the Josephson phase $\phi_n$ which is the 
difference of the wavefunction phases for each of the superconducting
electrodes that form the junction.
The flux induced by the qubit will be denoted as 
$\Phi_n^{(e)}={\Phi_0}\phi_n^{(e)}/(2\pi)$, $L$ is 
selfinductance of the JTL cell, $I_c$ is
the junction critical current and $\Phi_0$ is magnetic flux quantum.

The equations of motion for the Josephson phase can be obtained
in the standard way from the Hamiltonian (\ref{1}). They must
be complemented by the dissipative term that describes the
normal electron current $\hbar \dot{\phi}_n/(2eR)$ across the junction 
with $R$ being the
resistance of an individual junction. The external
dc bias $I_B$ is applied to each junction. Hence, the equations of
motion read
\begin{eqnarray}\label{2}
&&\frac{C\hbar}{2e}\ddot{\phi}_n+\frac{\hbar}{2eR}\dot{\phi}_n
+I_c\sin\phi_n-\\
&&-\frac{\Phi_0}{2\pi L}\hat{\Delta}\phi_n=I_B+\frac{\Phi_0}{2\pi L}
\left (\phi_n^{(e)}-\phi_{n-1}^{(e)}
\right), \nonumber \\
&& \nonumber \hat{\Delta}\phi_n \equiv \phi_{n+1}-2\phi_n+\phi_{n-1},\;\
n=\overline{1,N}.
\end{eqnarray}
Throughout the paper only the linear JTLs are to be described, therefore
the open ends boundary conditions will be used.
It is convenient to introduce the dimensionless variables
\begin{eqnarray}
&&H\to \frac{H}{E_J}; ~t\to t\omega_J,\omega_J=\sqrt{\frac{C\hbar}{2eI_c}} ;\\
\nonumber
&&~\gamma=\frac{I_B}{I_c};~\kappa=\frac{\Phi_0}{2\pi L I_c}.
\end{eqnarray}
As a result a dimensionless damped and dc driven 
discrete sine-Gordon (DSG) equation is obtained:
\begin{eqnarray}\label{DSG}
&&\ddot{\phi}_n+\alpha \dot{\phi}_n+\sin\phi_n
-\kappa \hat{\Delta}\phi_n=\\
\nonumber
&&=\gamma+f_n,~n=\overline{1,N},
\end{eqnarray}
where $\gamma$ denotes the dimensionless bias current and the constant
$\kappa$ is the measure of the discreteness. When $\kappa$ decreases the
discreteness effects become stronger. In the opposite limit $\kappa \to \infty$
they disappear and the continuum limit is restored.
The spatially inhomogeneous term $f_n$ is written in the following form
\begin{eqnarray}\nonumber
&&f_n=\frac{\Phi_0}{2\pi L I_c} \left (\phi_n^{(e)}-\phi_{n-1}^{(e)}\right)=\\
\nonumber
&&=
\frac{1}{LI_c} \left(\Phi_n^{(e)}-\Phi_{n-1}^{(e)} \right)=\\
&&=\sigma \kappa \left\{ \begin{array}{ccc}
0&,&n\neq n_0-1,~n_0, \\
|\mu| &,&n=n_0-1,\\
-|\mu| &,&n=n_0 .
\end{array}
\right. \label{f}
\end{eqnarray}
Here $\sigma=\mbox{sign} \mu=\pm 1$ is the qubit polarity, which defines
its state, 
the parameter $\kappa \mu$ measures the strength of the impurity and
equals 
\begin{equation}
\kappa \mu =\frac{\kappa}{\kappa_q}=\frac{L_q I_q}{ L I_c},\;\;
\kappa_q=\frac{\Phi_0}{2\pi L_q I_q},
\end{equation}
Here $I_q$ is the current in the qubit cirquit.
The total dimensionless energy reads
\begin{eqnarray} \label{ener1}
&&E=E_k+E_p,\;
E_k=\frac{1}{2}\sum_{n=1}^N \dot{\phi}^2_n,\\
&&E_p=\sum_{n=1}^N \left [\frac{\kappa(\phi_{n+1}-\phi_n)^2}{2}+
\right.\\
&& \left.+2\sin^2{\frac{\phi_n}{2}}+(\gamma +f_n)\phi_n \right].  
\nonumber
\end{eqnarray}
In the following sections we will analyze both analytically and 
numerically fluxon interaction with the qubit.

\section{Fluxon transmission through the qubit and the threshold current.}
\label{sec3}

\subsection{General remarks about topological soliton interactions with
impurities}

The problem of topological soliton interaction with spatial 
inhomogeneities in both the continuous and discrete systems 
has been studied thoroughly during the several last decades 
(see the respective chapters in the reviews and books 
\cite{km89rmp,gk92pr,bk03}). 
It is useful to recall several important
facts.  Generally speaking, solitons can get pinned by the
inhomogeneity whenever it is lattice discreteness or just
an individual impurity. If the system has dissipation and is biased by the 
constant current, the most typical situation consists of the following
scenarios:
\begin{itemize}
\item $0<\gamma < \gamma_{thr}$. In this case the
inhomogeneity is so strong that a soliton cannot pass it. The value 
$\gamma_{thr}$ is known as the threshold
current. It is the minimal current for which a soliton can propagate.
\item $\gamma_{thr} < \gamma < \gamma_c$. The bias is strong
enough to sustain moving solitons. Both moving and
standing (pinned by the inhomogeneity) solitons can exist.
Depending on the initial condition the system can settle on any
of these states.
\item $\gamma>\gamma_c$. Bias is so strong that impurities
cannot trap the soliton. 
\end{itemize}
Discreteness obstructs free soliton propagation along the array.
There exists a critical depinning current $\gamma_{PN}$ such
that for $\gamma < \gamma_{PN}$ no soliton propagation is
possible.
If $\kappa$ is too small the soliton cannot move.
We will consider only the values of $\kappa$ large enough that 
$\gamma_{PN}<\gamma_{thr}$.

\subsection{Manifestation of the discreteness effects }
\label{sec3.2}
\begin{figure}[htb]
\includegraphics[width=1.01\linewidth]{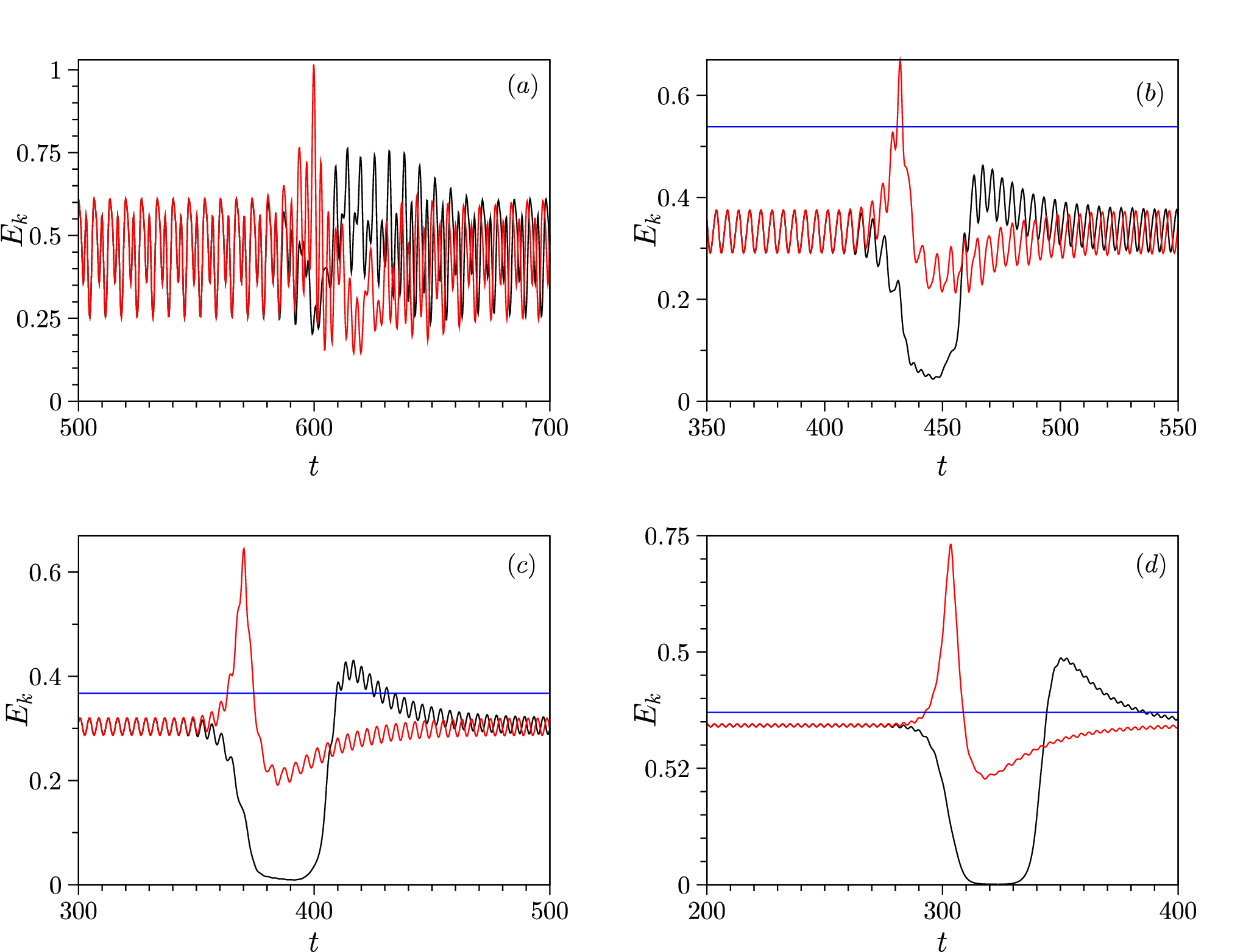}
\caption{(Colour online.) Total kinetic energy (\ref{ener1}) as a function of time
for the fluxon interacting with the qubit ($\sigma=1$, black and
$\sigma=-1$, red) for $\kappa=0.5$, $\gamma=0.0487$ (a), 
$\kappa=0.75$, $\gamma=0.0251$ (b), 
$\kappa=1$, $\gamma=0.0193$ (c) and $\kappa=1.5$, $\gamma=0.0175$ (d). 
Other parameters are: $\alpha=0.05$
and $|\mu|=0.2$. Solid blue line corresponds to the fluxon kinetic
energy in the continuum case $E_k=4\sqrt{\kappa} v^2_\infty $.}
\label{fig1}
\end{figure}

\begin{figure}[htb]
\includegraphics[width=.99\linewidth]{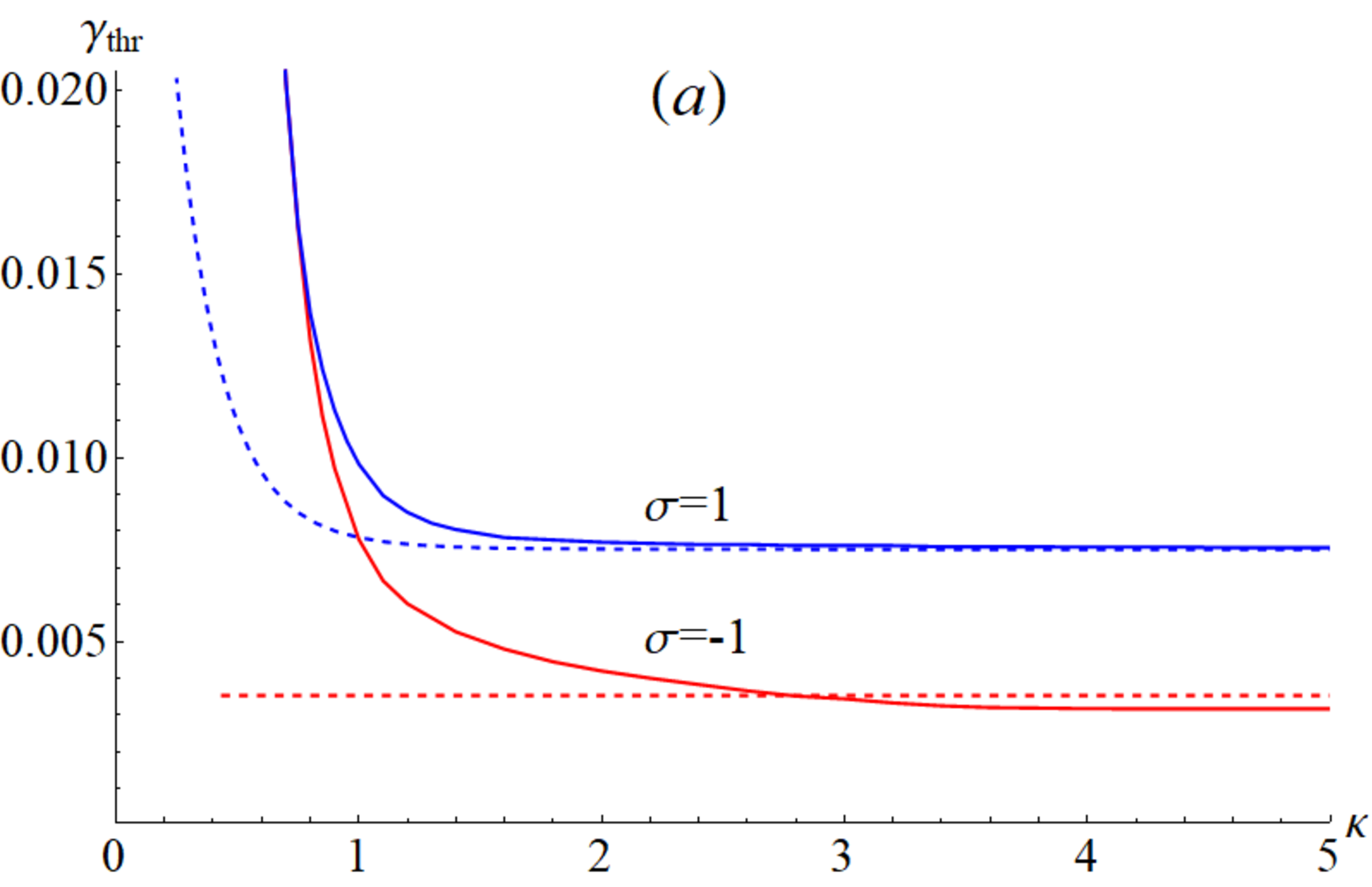}
\includegraphics[width=.99\linewidth]{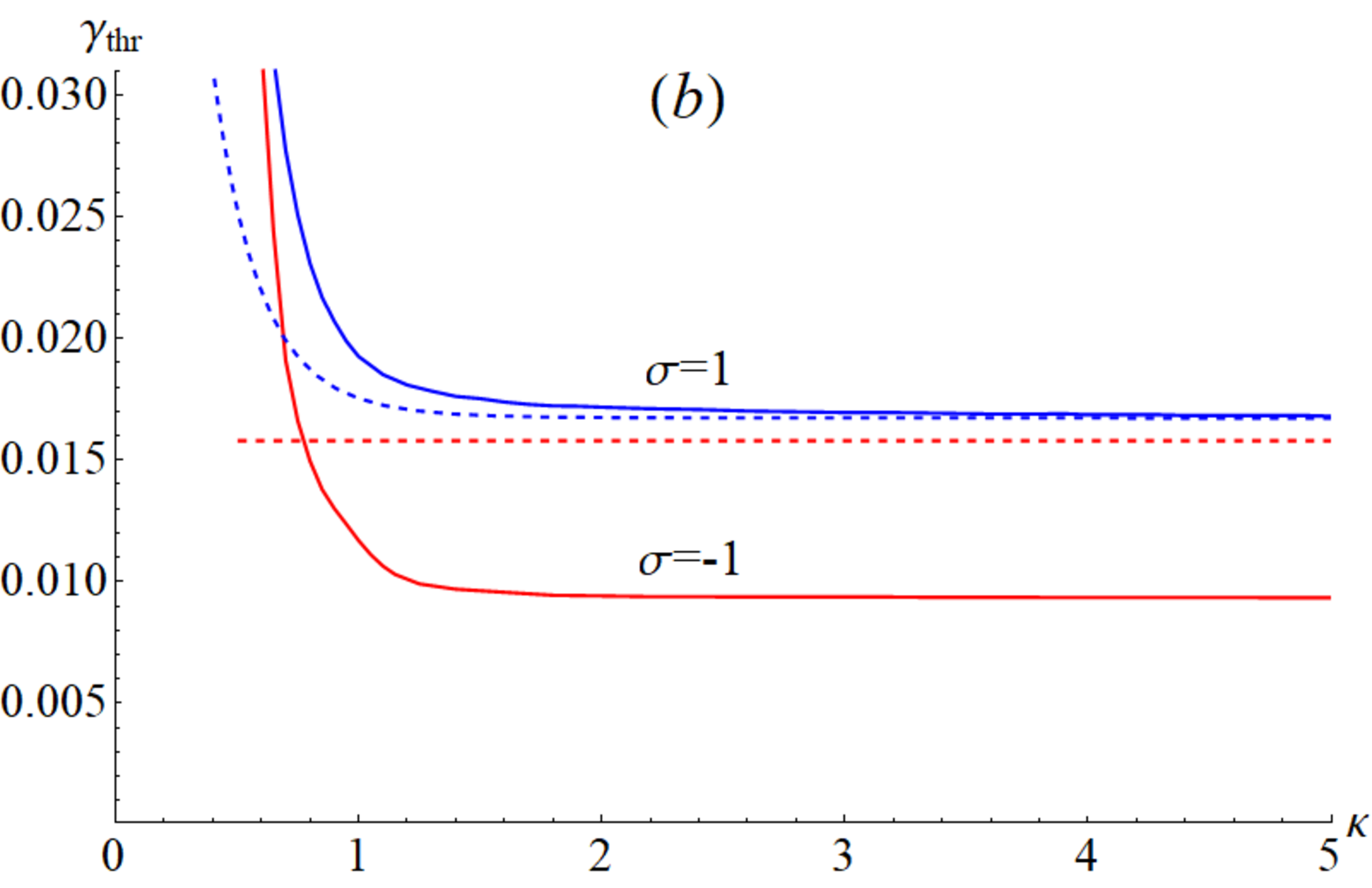}
\caption{(Colour online.) Threshold current as a function of the discreteness
parameter for $|\mu|=0.2$ and $\alpha=0.02$ (a) and $\alpha=0.05$ (b).
Red colour corresponds to $\sigma=-1$ and blue line is $\sigma=1$. 
Solid lines correspond to the numerically computed results, dashed line
is analytical approximation (see text for details). }
\label{fig2}
\end{figure}

\begin{figure}[htb]
\includegraphics[width=.49\linewidth]{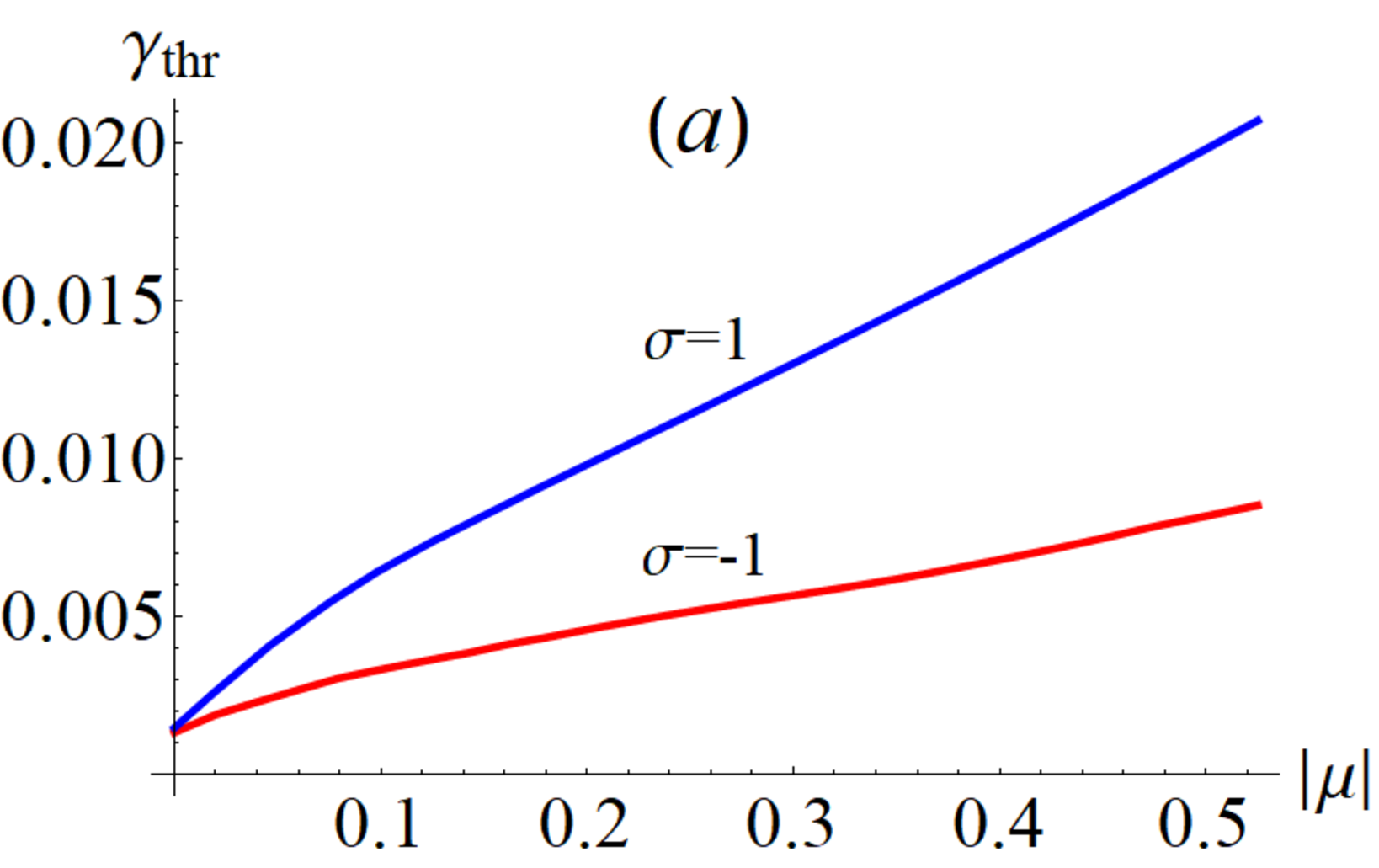}
\includegraphics[width=.49\linewidth]{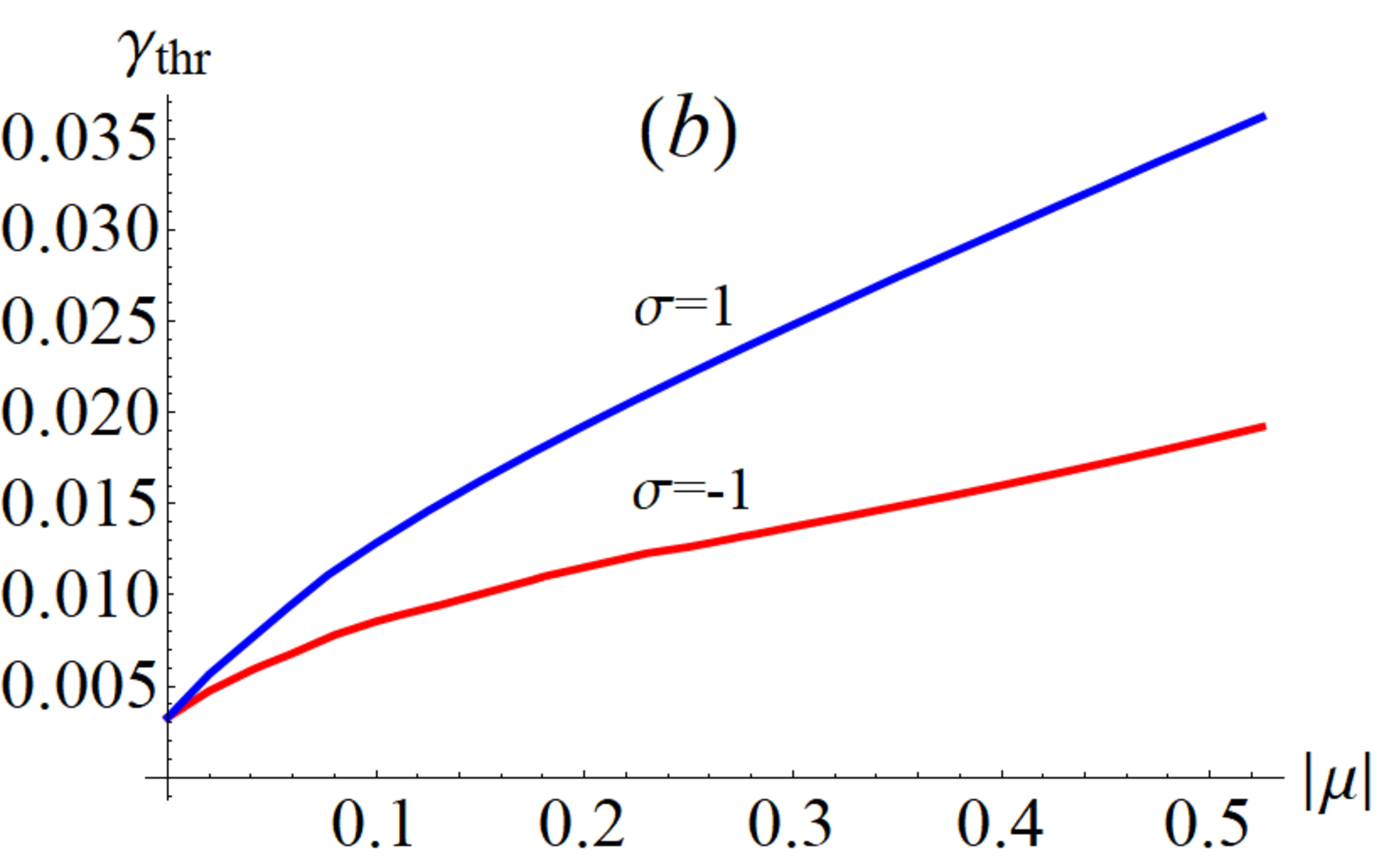}
\includegraphics[width=.49\linewidth]{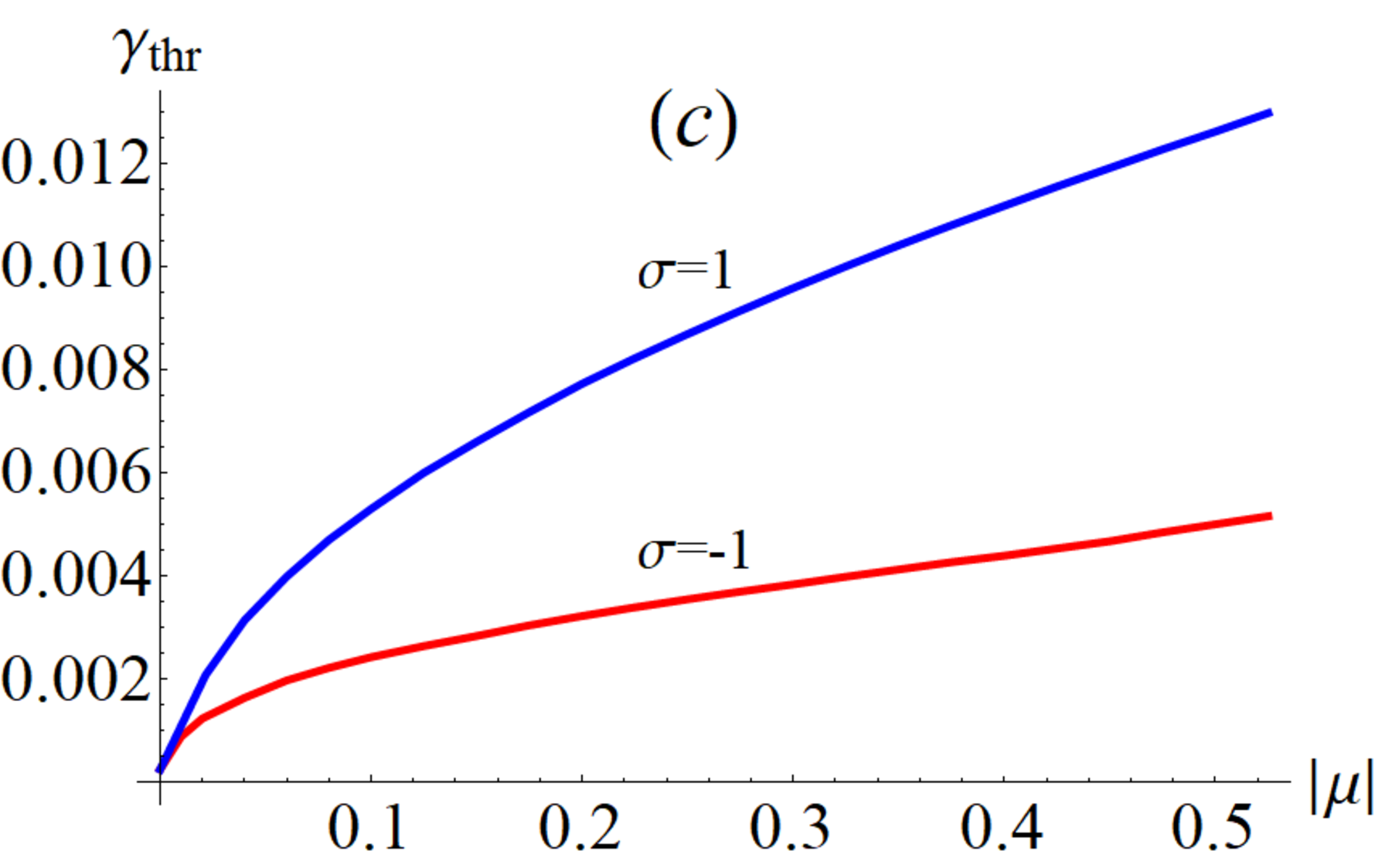}
\includegraphics[width=.49\linewidth]{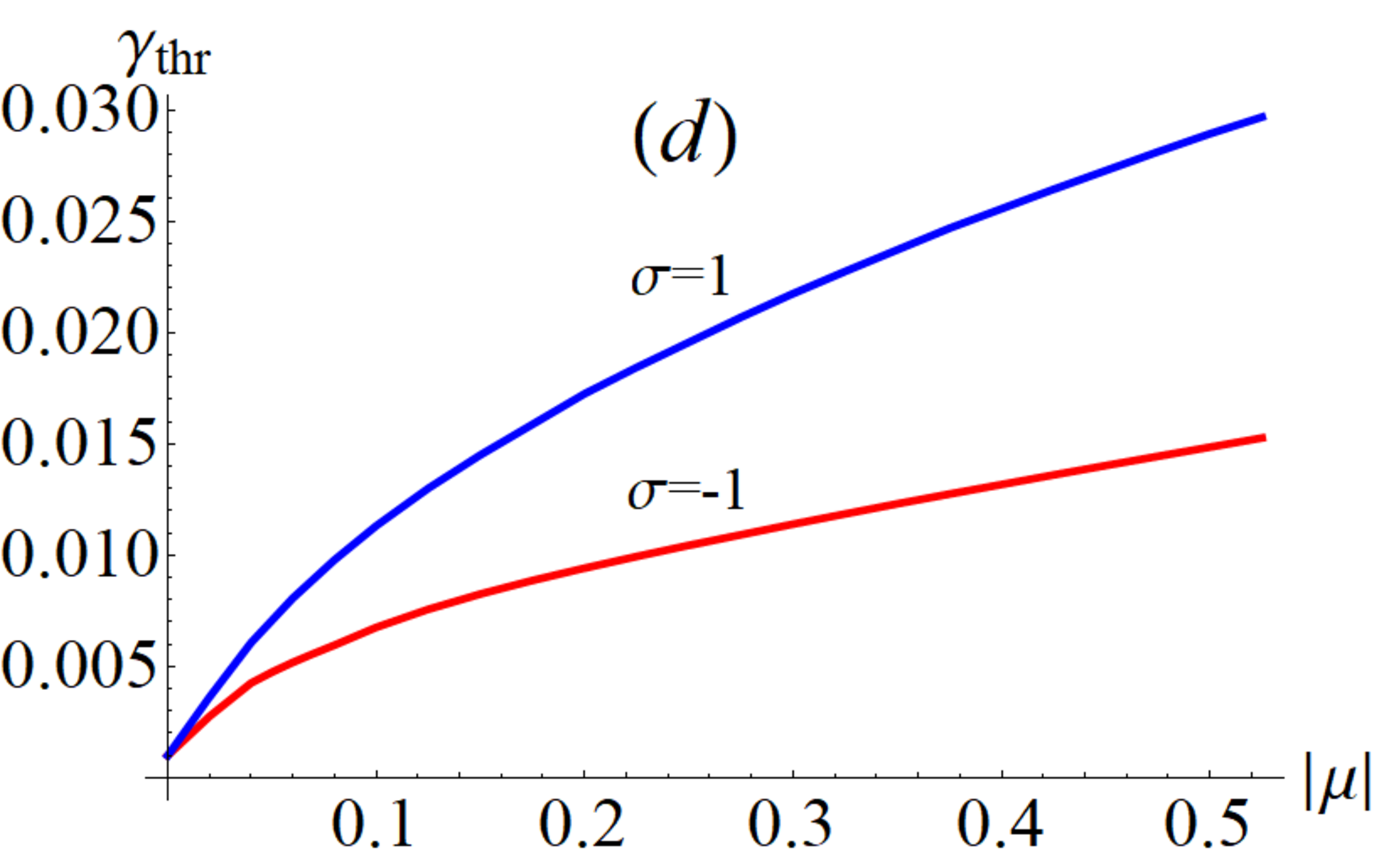}
\caption{(Colour online.) Threshold current as a function of the impurity
strength $\mu$  for $\kappa=1$: $\alpha=0.02$ (a) and $\alpha=0.05$ (b);
 $\kappa=2$: $\alpha=0.02$ (c) and $\alpha=0.05$ (d). Blue colour corresponds
 to $\sigma=1$ and red corresponds to $\sigma=-1$.}
\label{thr_mu}
\end{figure}

In this subsection we will compute numerically 
the threshold current $\gamma_{thr}^{\pm}=\gamma_{thr}(\sigma=\pm 1)$
for the impurities with the different polarities $\sigma$.
For the numerical simulations of the DSG equation (\ref{DSG}) the 4th
order Runge-Kutta method will be used.
The details of the fluxon interaction with the dipole impurity are
given in Fig. \ref{fig1}, the threshold current as a function of the
system parameters is given in Figs. \ref{fig2}-\ref{thr_mu}. 
In this subsection the JTL size is chosen to be large enough ($N=300\div 600$) to avoid any boundary effects. A special case of short JTLs is discussed separately in Subsec. \ref{ssec4.2}.

Fluxon transmission through a lattice inhomogeneity has certain
principal differences from the respective continuum case. This
is illustrated by the time dependencies of the total kinetic energy
of the JTL (\ref{ener1}) for the different values of the discreteness
constant $\kappa$ and external bias $\gamma$. The bias was chosen to
be slightly higher than the respective threshold value $\gamma^\pm_{thr}$.
These dependencies are presented in Fig. \ref{fig1}.
The fluxon dynamics in the vicinity of the qubit impurity depends on the
impurity polarity. For the positive polarity ($\sigma=1$), the fluxon
slows down before approaching the qubit and accelerates back to its
equilibrium velocity.
%
%
%
For the $\sigma=-1$ the situation is the opposite. The fluxon accelerates
first and then slows down back to the equilibrium velocity. This
is consistent with the continuum approximation \cite{sz19pla,ag84jetpl} 
in which
the $\sigma=1$ impurity is felt by the fluxon as a potential barrier. On the
contrary, the fluxon sees the $\sigma=-1$ impurity  as a potential well.
Another useful observation is the amplitude of the kinetic energy
oscillations. These oscillations appear due to discreteness and are rather 
weak even for the sufficiently discrete JTL with $\kappa=1.5$ (see Fig. \ref{fig1}d).
As the discreteness increases, these oscillations increase as well, 
eventually reaching the situation when they are of the order of the average kinetic energy value (see Fig. \ref{fig1}a). This means that the
contribution of the oscillations around the fluxon center of mass is very strong.

If discreteness is weak ($\kappa \gg 1$), it is possible to project
the many-body problem described by the DSG equation (\ref{DSG}) onto the
two-dimensional subspace of the collective coordinates $(X,v)$ where
$X=X(t)$ it the fluxon center of mass and $v(t)$ is its velocity.
The discreteness of the media is modelled by the Peierls-Nabarro (PN) 
potential. Using the technique described in \cite{bk91prb,bk03} one 
arrives to
the collective equations of motion on the fluxon center of mass $X$,
$\phi(x,t) \approx 4 \tan^{-1} \exp {[x-X(t)]}$, $x\approx nh$, $h=1/\sqrt{\kappa}$:
\begin{eqnarray}\nonumber
&&8\ddot{X}+8\alpha \dot{X}=-\frac{\partial U}{\partial X}= \\
&&=-\frac{\partial}{\partial X}\left[U_{0}(X)+U_\gamma(X)+U_{PN}(X) \right].
\label{7}
\end{eqnarray}
This is a newtonian equation of motion for the dissipative 
particle with the mass $m=8$ in the field created by the potential $U(X)$.
This potential consists of three parts. The first one, $U_0(X)$ is
created by the impurity, the second one, $U_\gamma(X)$, appears due to the
external bias. The last one is the PN potential $U_{PN}(X)$.
The explicit form of these potentials is written as follows:
\begin{eqnarray}\label{10}
&&U_{PN}(X)=\Delta_{PN} \cos (2\pi \sqrt{\kappa}X), \\
&& \Delta_{PN} =\frac{16\pi^2\sqrt{\kappa}}{\sinh(\pi^2\sqrt{\kappa})}, \label{10a}\\
&&U_0(X)=\frac{2\mu\sigma}{\cosh{X}},\label{11}\\
&&U_\gamma (X)=
-2\pi \gamma X-\frac{2\gamma}{\sqrt{\kappa}} \frac{\sin(2\pi X\sqrt{\kappa})}{\cosh(\pi^2\sqrt{\kappa})}.
\label{12}
\end{eqnarray}
It appears that the potential $U_0(X)$ does not depend on discreteness. 

In the continuum limit ($x\approx nh$) the DSG equation 
(\ref{DSG}-\ref{f})
transforms into continuous sine-Gordon (SG) equation
\begin{eqnarray}\label{SG}
&&\phi_{tt}- \phi_{xx}+\sin\phi+\alpha \phi_t=\\
&&=\sigma |\mu| \delta'(x)+\gamma, \; \nonumber
\end{eqnarray}
which has been studied previously \cite{fssk-s07prb,sz19pla,ag84jetpl}.
In the continuum limit the 
collective-coordinate equations (\ref{7})-(\ref{12}) 
will contain only the potentials $U_0(X)$ and $U_\gamma(X)=-2\pi\gamma X$.
The PN potential will naturally be absent.
As a result, the approach
developed in papers \cite{sz19pla,kmn88jetp} can be applied. 
Thus, the threshold currents are given by (see Ref. \cite{sz19pla} for
details):
\begin{eqnarray}\nonumber
&&\gamma_{thr,c}^{+}={\Upsilon}^+_1\alpha+\Upsilon_2^+\alpha^2=\\
\label{14}
&&=\frac{2\sqrt{{2\mu}}}{\pi}\alpha -\frac{8}{\pi}
\ln \left(\frac{1+\sqrt{2}}{\sqrt{2}}\right)\alpha^2~,\\
\nonumber
&&\gamma_{thr,c}^{-}=\frac{4\Gamma \left (\frac{1}{4}  \right)}{\pi^{5/4}}
 \left |{\mu} \right  |^{1/4}\alpha^{3/2}+\\
&&+ 12\frac{\alpha^2}{\pi}
\ln \left [\left ( \frac{2\Gamma \left (\frac{1}{4}  \right)}{\pi^{1/4}} \right )^{2/3} \frac{\alpha}{\sqrt{|\mu|}}\right]~. \label{15}
\end{eqnarray}
In the positive polarity case the first term $\Upsilon_1^+ \alpha$
can be obtained from the kinematic approach \cite{ms78pra}. The
second term $\Upsilon^+_2 \alpha^2$ is a correction that requires more elaborate approximation
\cite{kmn88jetp}. The ${\cal O}(\alpha)$ term is sufficient to have the
crudest approximation. The kinematic approximation
states that the threshold current is found
from the condition that the total fluxon kinetic energy is spent
to overcome the qubit-created potential. For the positive polarity $\sigma=1$
 this means $E_k=4v^2_\infty = U(X_{max})-U(X_{min})$. If we assume that
the incoming fluxon velocity is small and does not differ significantly
from its continuum value, we can use $v_\infty \approx \pi\gamma/(4\alpha)$.
The potential barrier $U(X_{max})-U(X_{min})$ must contain two contributions: the 
height of the impurity potential $2\mu$ [see Eq. (\ref{11})] and the
height of the PN potential $2\Delta_{PN}$ [see Eq. (\ref{12})]. As
a result we obtain the first order approximation to the threshold current in JTL 
\begin{equation}\label{16}
\gamma^+_{thr,1}=\frac{4\alpha}{\pi}\sqrt{\frac{\mu+\Delta_{PN}}{2}},
\end{equation}
where the PN barrier amplitude $\Delta_{PN}$ is given in Eq. (\ref{10a}).
In Fig. \ref{fig2} the numerically calculated dependencies $\gamma^{\pm}_{thr}(\kappa)$
are shown. Solid lines correspond to the numerically computed data. 
Dashed lines represent analytical approximations. 
In the continuum limit ($\kappa \to \infty$) the dependencies for 
both qubit polarities become flat. 
%
%
%
The threshold currents for both $\sigma=1$ and $\sigma=-1$ qubits 
increase strongly with the increase of discreteness. This is
quite natural because the fluxon needs more energy to overcome not
only the impurity potential but the PN potential as well.
Another important point is that the difference $\gamma^+_{thr}-\gamma^-_{thr}$
decreases as $\kappa \to 0$. 
This can be explained by the fact that
the obstructive influence of discreteness is so strong that it does
not discriminate between different polarities of the qubit. In other words,
the influence of the PN potential is of the same order as of the impurity
potential. This situation is illustrated quite clearly in Fig. \ref{fig1}d,
where the time oscillations of the kinetic energy due to the PN barrier 
are of the same order as the kinetic energy itself.
At some value of $\kappa$ the dependencies
coincide, however, we did not follow them to very small values because
of difficulty in defining $\gamma_{thr}$. This point will be discussed below.
For the $\sigma=+1$ case the continuum limit
(\ref{14}) is restored with high accuracy. In order to compare our
numerical results with the continuum limit, we have plotted the combined
approximation $\gamma^+_{thr}(\alpha)=\gamma^+_{thr,1}(\alpha)+\Upsilon_2^+ \alpha^2$ where
the first term comes from the kinematic approximation (\ref{16}) and
the second one is the second term from the continuum limit correction (\ref{14}). The abovementioned analytical approximation
captures the main features of the $\gamma_{thr}(\kappa)$ dependence but
is not that accurate for $\kappa \sim 1$. This means that the collective-
coordinate PN approximation (\ref{7})-(\ref{12}) does not work well for strongly discrete
lattices. In the case of $\sigma=-1$ we have plotted only the continuum
approximation (dashed red line) of the threshold current (\ref{15})
because the analytical approximation appears to be highly complicated.
One can observe that the coincidence between the numerical and analytical
results in the continuum limit varies. The answer lies in the fact that
the validity of the approximation, which allows to obtain Eq. (\ref{16}),
is limited. For details one can consult Ref. \cite{sz19pla}. 
The approximation (\ref{16}) is obtained under the assumptions 
$\alpha \ll |\mu| \ll 1 $, $\gamma \ll |\mu|\ll 1$, and 
$\alpha \ll \sqrt{|\mu|/2} $. In the
case $\alpha=0.05$ the last inequality is fulfilled in a rather weak sense:
$\alpha =0.05 < 0.33$. In the case $\alpha=0.02$ it works much better.
We conclude that the threshold current difference should be increased
if one wishes to increase the sensitivity of the qubit readout process. 
We observe that discreteness reduces this interval, hence, it plays
a destructive role in the qubit readout. 

Next, one can look at the dependence of the threshold current as
a function of the impurity strength. These dependencies are given
in Fig. \ref{thr_mu}a-d. Qualitatively, the power law ($\mu^{1/2}$ for $\sigma=1$ 
and $|\mu|^{1/4}$ for $\sigma=-1$)
is preserved.  
%
%
%
The difference $|\gamma^+_{thr}-\gamma^-_{thr}|$ increases as $\kappa$
increases and decreases as $\alpha$ decreases.
The only principal difference can be seen in the limit $\mu\to 0$. The
threshold current does not vanish and this is a consequence of
the discreteness. Even if there is no impurity some non-zero bias
is necessary to overcome the PN barrier. In the continuum the threshold current would be exactly zero for $\mu=0$.

\begin{figure}[htb]
\includegraphics[width=1.03\linewidth]{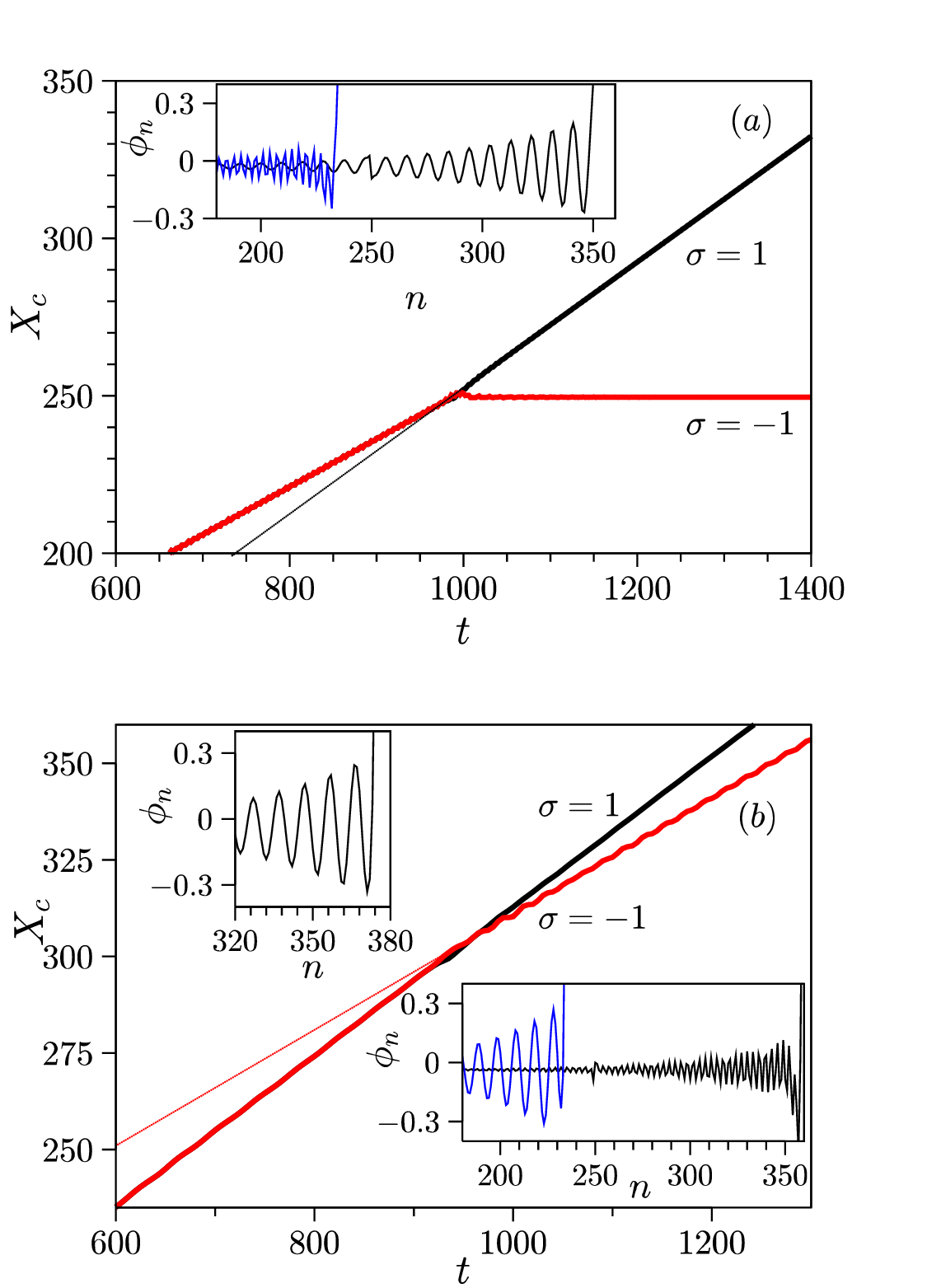}
\caption{(Colour online.) Fluxon center of mass evolution for $\alpha=0.02$, 
$\kappa=0.6$, $\gamma=0.0285$ (a) and $\kappa=0.55$, $\gamma=0.037$ (b).
Thick black and red lines correspond to the positive and negative
qubit polarity, respectively. Thin dotted straight lines  
satisfy $X_c \propto v t$,
where $v$ is the fluxon velocity after interaction with the qubit. 
They are plotted to highlight the fluxon velocity change
after the interaction.  
The insets show the details of the fluxon tail before (blue)
and after (black) the interaction. See text for details.}
\label{fig4}
\end{figure}

Calculation of the fluxon scattering in the limit of very strong
discreteness is complicated due to different reasons. Among them there
are dynamical chaos, creation of fluxon-antifluxon pairs and 
 the multiplicity of the moving dc-driven
fluxons that coexist for the same value of the dc bias. 
It is well known for the finite size JTLs \cite{ucm93prb,wzso96pd,p08prb}  
that the discreteness of the media 
brings the fundamental change to the velocity-bias $v=v(\gamma)$
dependence, which switches from the monotonous curve in the continuum
to the piecewise dependence in the discrete case.
Each curve corresponds to the resonance with the particular cavity mode.
The hysteresis is clearly seen for the large bias values while
for the small bias and, especially for the weak discreteness the $v=v(\gamma)$
it hardly can be observed. In the almost infinite JTLs the cavity modes 
play no role because they decay over the lattice. 
However,  we have observed how the
coexistence of two fluxons with the different essentially non-relativistic 
velocities 
can be manifested. In Fig. \ref{fig4} the change of the 
fluxon velocity after the scattering is demonstrated for rather
small values of discreteness, namely $\kappa=0.55$ and $\kappa=0.6$.
The fluxon is launched from the same starting point, therefore
both the trajectories coincide before the interaction with the qubit.
But they diverge after the interaction.
%
%
%
In the case shown in Fig. \ref{fig4}a the fluxon attains higher velocity
after passing the qubit with $\sigma=1$. After passing the $\sigma=-1$
qubit the fluxon gets trapped.
The lattice size is $N=500$, the integration time
is $t \sim 1000 \gg 1/\alpha$. So it is quite clear that before and
after passing the impurity we are dealing
with the different attractors of the system that have different 
equilibrium velocities. Before hitting the impurity the fluxon had velocity
$v\approx 0.15$ and after the interaction it acquired higher
velocity $v\approx 0.20$. The thin dashed line $X_c \propto 0.2 t$ is plotted
in order to highlight the velocity change. 
The oscillating tails behind
the fluxon additionally confirm  that these are two different fluxon
attractors that are coupled to plasmons with different wavelengths (see
the respective inset). This is typical situation \cite{bhz00pre}, when the topological solitons, moving with different velocities, excite different plane waves with different wavelengths. These wavelengths satisfy the resonance condition
$vq=\omega_L(q)$, where $\omega_L(q)=[(1-\gamma^2)^{1/2}+4\kappa \sin^2(q/2)]^{1/2}$ is the plasmon dispersion law that
can be easily derived from the DSG equation. This equation always has
at least one root and if $v$ is small enough it will have multiple roots.
A similar situation is illustrated in Fig. \ref{fig4}b.
Here the fluxon passes the $\sigma=1$ qubit without any change of velocity.
Also, the wavelength of the plasmon tail remains the same, as can
be observed after comparing the upper inset (after interaction) 
and the blue dependence
in the lower inset (before interaction). On the other hand, the fluxon
loses its velocity after passing the $\sigma=-1$ qubit. Comparison of the
oscillating tails behind the fluxon before and after the scattering 
clearly shows the change in the tail wavelength.

\section{Qubit delay time}
\label{sec4}

\subsection{Qubit delay time for the infinite JTL.}
\label{ssec4.1}
In the fluxon assisted qubit readout process it is important
to maximize the time difference between the fluxon passage
time through the $\sigma=-1$ and $\sigma=+1$ impurities. 
By measuring those times, an experimentalist can determine
the qubit state. Thus, we have focused on the computation
of the difference  
\begin{equation}
\Delta t=|t_+-t_-|,
\label{18}
\end{equation}
which we should call the delay time and 
where $t_\pm$ is
the time necessary for the fluxon to travel from one end of
the JTL to another one. The subscripts in $t_\pm$ correspond
to the respective polarity of the impurity, $\mbox{sign} (\sigma)$.

\begin{figure}[htb]
\includegraphics[width=1.\linewidth]{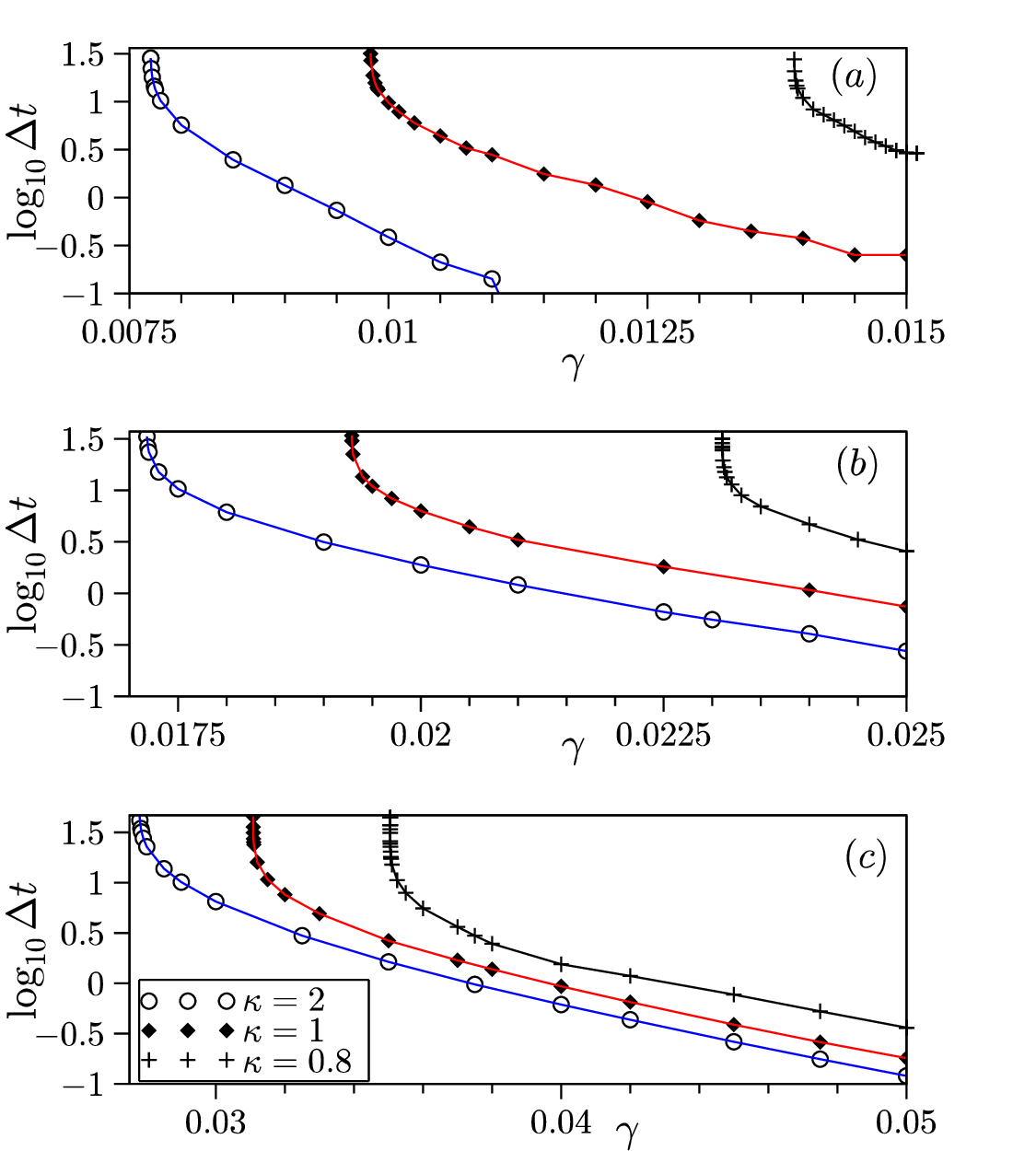}
\caption{(Colour online.) Decimal logarithm of the delay time on the qubit as a function of the 
dimensionless
bias for the different values of the discreteness parameter and damping 
[$\alpha=0.02$ (a), $\alpha=0.05$ (b) and $\alpha=0.1$ (c)]. The discreteness 
values are given
in the legend and $|\mu|=0.2$. Solid lines are drawn as a guide for an eye.}
\label{fig5}
\end{figure}

The main results are given in Fig. \ref{fig5} where the 
dependence of the delay time is given as a function of the external
bias $\gamma$ for the different values of the 
JTL discreteness parameter $\kappa$. 
%
%
%
One can observe that the delay time increases significantly for the
fixed bias value as the JTL
becomes more discrete. For example, compare the values of $\Delta t$
for the same values of damping ($\alpha=0.05$) and bias ($\gamma=0.02$).
For $\kappa=1$ we have $\Delta t \approx 2.13$ and for $\kappa=2$ we have
$\Delta t \approx 6.3$, hence a threefold increase when the coupling constant
is decreased by  half.
The singular behavior of $\Delta t$ at some
bias value can be clearly seen. This value equals $\gamma_{thr}^+$ at which the fluxon
still has enough energy to pass the $\sigma=-1$ qubit, 
but is trapped by the $\sigma=1$ qubit. 
Thus, the bias 
values $\gamma\gtrsim \gamma_{thr}^{+}$ are the most desirable from
the point of view of best readout sensitivity.

\subsection{Influence of the JTL size on the delay time}
\label{ssec4.2}

\begin{figure}[htb]
\includegraphics[scale=0.45]{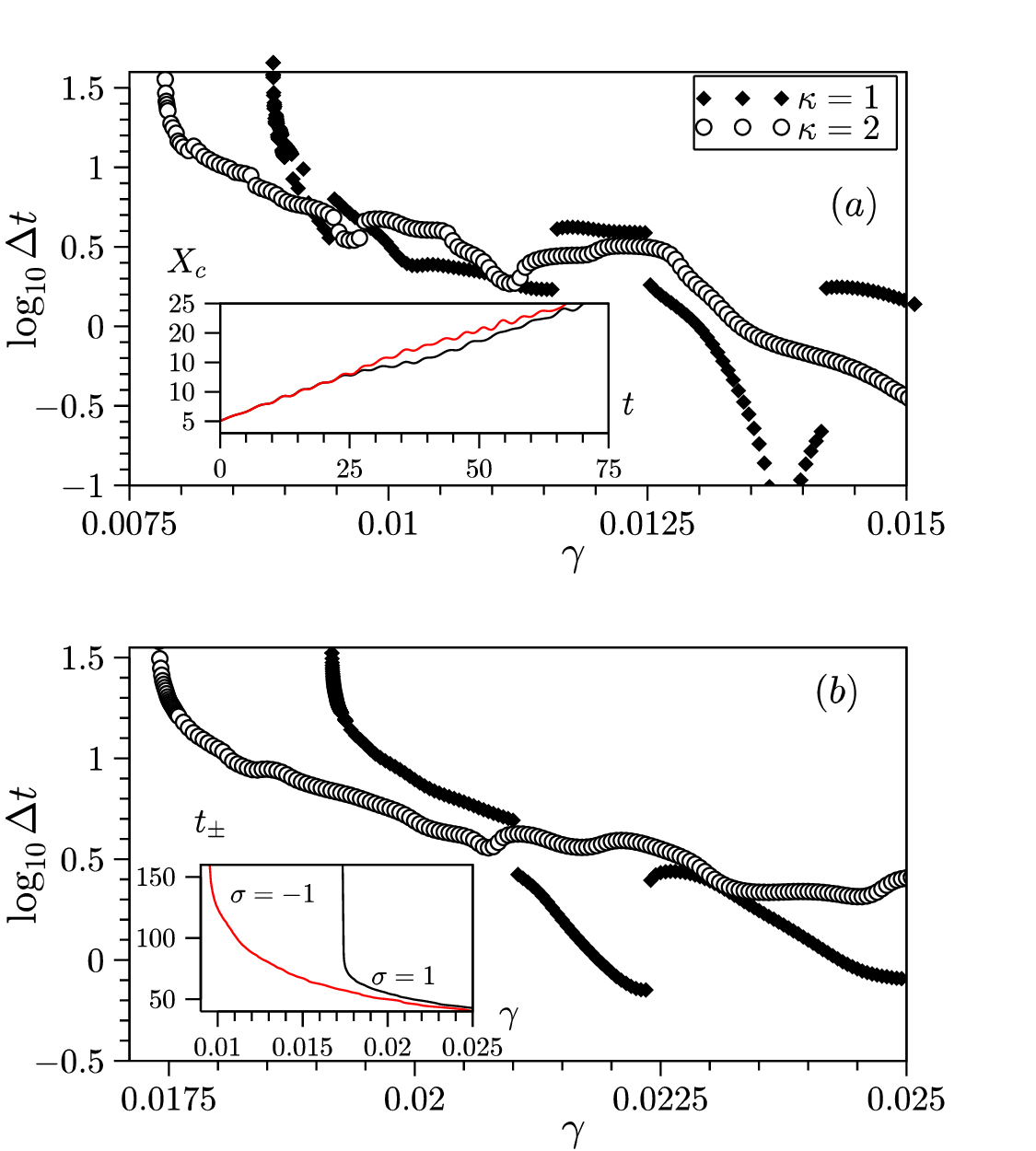}
\caption{(Colour online.) Decimal logarithm of the delay time for the array with $N=30$ junctions 
as a function of the bias $\gamma$ for $\kappa=1$ and $\kappa=2$ (see
the legend)  and $\alpha=0.02$ (a), $\alpha=0.05$ (b).  
The inset in panel (a) shows the trajectories of the fluxon
for interacting with the $\sigma=1$ (black line) and $\sigma=-1$ (red line) 
impurities. The dc bias equals $\gamma=0.01$. The inset in panel (b) 
the passage $t_\pm$ times
are given for the case $\kappa=2$, $\alpha=0.05$.  
In all figures $|\mu|=0.2$.}
\label{fig6}
\end{figure}

In most of the experiments with the JTLs their length is not very
big. They are significantly shorter than used in the numerical 
simulations in the previous sections. Normally the array length is
about $N \sim 10 \div 30$ junctions \cite{tmo00prl,baufz00prl,p08prb}. In 
such short arrays the boundary effects are important. In order to
assess them the simulations for the short linear JTL with $N=30$ 
have been performed. 
We simulated numerically the situation when the soliton was
launched at the site $n=5$ and computed the passage time that
elapsed when the fluxon reached the site $n=25$.
The delay time $\Delta t$ as a function of the
external bias $\gamma$ is given in Fig. \ref{fig6}. Different 
panels correspond to the different values of damping $\alpha$.
One can observe the main features from the
almost infinite junction case discussed in the previous subsection.
Strongly non-monotonic  behaviour appears because the fluxon center
of mass has an oscillating component. If dissipation is too small,
as in the case of $\alpha=0.02$ (Fig.\ref{fig6}a), these oscillations
still persist when the fluxon reaches the end of the junction, as 
shown in the inset to that figure. Thus, if the bias is too strong, the
difference between the passage times for the different qubit states is
less than the oscillation of the fluxon center. Therefore, no 
qubit readout is possible. However, for the bias values 
$\gamma\gtrsim \gamma^+_{thr}$ the delay time is well defined similarly to
the infinite JTL case described in Subsec. \ref{ssec4.1}.
The delay time $\Delta t$ for the fixed bias increases as $\kappa$
is decreased. For the larger values of dissipation the dependencies become
more regular because the oscillating component of the fluxon center
of mass decays faster. The threshold values $\gamma^{\pm}_{thr}$ are
approximately the same as in the almost infinite case (see the inset to
Fig. \ref{fig6}b).

\section{Discussion and conclusions}
\label{disc}

This article is devoted to the studies of the fluxon interaction
with the dipole impurity in the Josephson junction array [Josephson
transmission line (JTL)]. 
The dipole impurity describes the inductively coupled to the JTL
qubit. The impurity has two values of polarity $\sigma=\pm 1$ that correspond
to different states of the qubit. By scattering a fluxon on the
qubit and measuring its passage time, one can distinguish between the
different qubit states because the fluxon interacts differently with
the $\sigma=-1$ and $\sigma=1$ impurities \cite{ars06prb,fssk-s07prb}.
Hence, it is important to maximize the difference $\Delta t$ between the abovementioned
passage times. In this work we addressed the question how the discreteness of the JTL influences $\Delta t$.

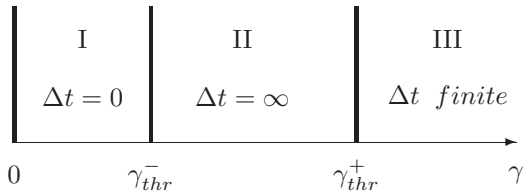
\begin{figure}[htb]
\setlength{\unitlength}{6.cm}
\begin{picture}(1.9,0.45)
\linethickness{0.005cm}
\put(0.2, 0.1){\vector(1,0){1.1}}
\linethickness{0.05cm}
\put(0.2, 0.1){\line(0,1){0.3}}
\put(0.5, 0.1){\line(0,1){0.3}}
\put(0.95, 0.1){\line(0,1){0.3}}
\put(0.2, 0.03){\makebox(0,0){$0$}}
\put(1.3, 0.03){\makebox(0,0){$\gamma$}}
\put(0.95, 0.03){\makebox(0,0){$\gamma^+_{thr}$}}
\put(0.5, 0.03){\makebox(0,0){$\gamma^-_{thr}$}}
\put(0.35, 0.2){\makebox(0,0){$\Delta t=0$}}
\put(0.35, 0.33){\makebox(0,0){I}}
\put(0.7, 0.2){\makebox(0,0){$\Delta t=\infty$}}
\put(0.7, 0.33){\makebox(0,0){II}}
\put(1.15, 0.2){\makebox(0,0){$\Delta t \;\;finite$}}
\put(1.15, 0.33){\makebox(0,0){III}}
\end{picture}
\caption{Schematic representation of the different scenarios of
the delay time detection as a function of the external bias.
(see text).}
\label{fig7}
\end{figure}

Our findings can be summarized by the diagram shown in Fig. \ref{fig7}.
Depending on the value of the external dc bias, there are three
sectors:
\begin{itemize}
\item[I.] $0<\gamma<\gamma^-_{thr}$. In this sector the bias is too small 
and fluxon propagation is not possible. Hence, no readout process.
\item [II.] $\gamma^-_{thr}<\gamma<\gamma^+_{thr}$. Fluxon
is always trapped on the $\sigma=+1$ qubit and, at the same time,
passes the $\sigma=-1$ qubit. Thus, the delay time is infinite. In this
case it is desirable to increase this bias interval. Our results demonstrate
that with the increase of discreteness this interval decreases. 
At some point ($\kappa \sim 0.5$) 
the threshold values for different polarities coincide. 
We did not venture for lower values of $\kappa$.
Thus, we conclude that discreteness poses a destructive effect for the
fluxon readout in this bias range.  
\item [III.] $\gamma>\gamma^+_{thr}$. In this case the fluxon passes
throughout the qubits with both polarities. The delay time
$\Delta t$  
helps us to assess the effectiveness of the readout process. It is
desirable to maximize this time. We have observed that the delay time 
increases 
significantly if discreteness becomes stronger. In particular, for the
fixed bias values, $\Delta t$ can increase by the factor of $2$ or even more
if the discreteness constant is halved. 
If discreteness slows down the fluxon motion it will naturally increase the
delay time.
Hence, if one applies the dc
bias at $\gamma \gtrsim \gamma^+_{thr}$ the sensitivity of the readout process
is better for the more discrete JTLs.
\end{itemize}

Also we would like to note that for the essentially discrete JTLs the
hysteresis effects on the velocity-bias (or current-voltage characteristics) can be strong. This means that for
the fixed bias value several fluxon attractors with different 
velocities coexist. As a result, a clear definition of the delay time on the
qubit is complicated because it depends on the initial conditions.
This hysteresis 
phenomenon is well-documented experimentally and theoretically 
\cite{ucm93prb,wzso96pd,p08prb} for the short junction when the
cavity modes play an important role.
In our work we have encountered yet another manifestation of this
hysteresis, but for the almost infinite lines, 
when the fluxon jumps to an attractor with a different velocity
after interacting with the qubit. This scattering can be both accelerating
and decelerating, e.g. the fluxon switches to the larger or smaller velocity
after the scattering depending on the qubit polarity. 
We believe that this phenomenon must occur not just for the
dipole impurity but for other types of impurities.
It requires a more detailed investigation that will be reported elsewhere.

\section*{Acknowledgements}
We would like to thank the Armed Forces of Ukraine for providing security
to perform this work.
Both the authors acknowledge a support by the National
Research Foundation of Ukraine grant (2020.02/0051)
"Topological phases of matter and excitations in Dirac
materials, Josephson junctions and magnets".


\end{document}